\documentclass[12pt]{article}
\usepackage{hyperref}
\usepackage{lscape}
\usepackage{amsmath}
\usepackage{amssymb}

\renewcommand{\title}[1]{%
    \bigskip%
    \begin{center}%
    \Large\bf #1%
    \end{center}%
    \vskip .2in}

\renewcommand{\author}[1]{%
    {\begin{center}
    #1
    \end{center}}}
\newcommand{\address}[1]{\vspace{-1.7em}\vspace{0pt}
    {\begin{center}
    \it #1
    \end{center}}}

\begin{document}

%\twocolumn[
 % \begin{@twocolumnfalse}

\title{Torsional Newton-Cartan geometry from Galilean gauge theory}

\author
{
Rabin Banerjee  $\,^{\rm a,b}$,
%Arpita Mitra    $\,^{\rm a, e}$,
Pradip Mukherjee $\,^{\rm c,d}$}
 %\footnote{Also, Visiting Associate, S. N. Bose National Centre for Basic Sciences, JD Block, Sector III, Salt Lake City, Kolkata -700 098, India  } 
%,
\address{$^{\rm a}$S. N. Bose National Centre 
for Basic Sciences, JD Block, Sector III, Salt Lake City, Kolkata -700 098, India }

\address{$^{\rm c}$Department of Physics, Barasat Government College,\\10, KNC Road, Barasat, Kol 700124.

 }

\address{$^{\rm b}$\tt rabin@bose.res.in}
\address{$^{\rm d}$\tt mukhpradip@gmail.com}
%\address{$^{\rm e}$\tt arpita12t@bose.res.in}

\begin{abstract}
Using the recently advanced Galilean gauge theory (GGT) we give a comprehensive construction of torsional Newton Cartan geometry. The coupling of a Galilean symmetric model with background NC geometry following GGT is illustrated by a free nonrelativistic scalar field theory. The issue of spatial diffeomorphisn \cite{SW, BMM3} is focussed from a new angle. The expression of the torsionful connection is worked out which is in complete parallel with the relativistic theory. Also 
smooth transition of the connection to its well known torsionless expression is demonstrated. A complete (implicit) expression of the torsion tensor for the Newton Cartan spacetime is provided where the first order variables occur in a suggestive way. The well known result for the temporal part of torsion is reproduced from our expression. 
\end{abstract}

%\tableofcontents
\section{Introduction}
%\begin{enumerate}
%\item
%{\bf{Introduction}}:
 Newton Cartan (NC) geometry \cite{Cartan-1923,Cartan-1924,Havas, ANDE,TrautA,Kuch,Daut,EHL, MALA} is a resurgent area of interest in the literature \cite{ABP, ABP1, Son, BMM2, BM4, X}, mainly because of its applications in various crucial topics such as % holographic theories of fluids \cite{R2},
  fractional quantum Hall effect (FQHE) \cite{BM4,SW,F, HS,M1,M2,Can,GS,Wu}, Horava - Lifshitz
gravity \cite{BM4, C1, A}, Galilean anomalies and hydrodynamics \cite{B1} etc. Inspite of its ubiquity there exists diverse and in some cases conflicting opinions about the details of NC geometry in the current literature. To cite an instance it will be sufficient to mention the issue of torsion. It appears from \cite{ABP, ABP1} that it is not possible to obtain torsional NC
geometry without including conformal symmetry in the usual gauging of the centrally extended Galilean algebra. However in \cite{C1}, where similar techniques have been employed , the existence of torsional NC geometry has been reported without the necessity of including conformal symmetry. This ambiguity 
is not negotiable as torsion is a  property of the NC geometry and its existence or otherwise should not depend on whether there is conformal symmetry or not.
% Another point of discord is the expression of the torsionful connection. Torsion is by definition the antisymmetric part of the connection. Full expression of the torsional connection is given in \cite{GS,J}. Inspite of their claim, the corresponding connection fail to retain the metric compatibility. There are reasons, however, for some confusion because the very nature of introduction and development of non relativistic diffeomorphism poses difficulties to introduce torsion.
Another related issue concerns the expression of the torsion in terms of the elements of NC geometry. Many authors give the 'temporal' part only. Full expressions for the torsional connection had been given in \cite{ABP1, GS} but they do not agree. In the latter, torsion is purely temporal and the expression for torsional connection may be criticised from several angles. The central issue is that torsional NC geometry should be viewed as the nonrelstivistic limit of Riemann Cartan spacetime. In most of the existing analysis, this point seems to be obscured.

  In fact, the concept of  nonrelativistic diffeomorphism invariance (NRDI), which led to the recent understanding of the NC spacetime, was marred with several problems, of which passage to Galilean symmetry in the flat limit is a representative one. We have shown that this problem disappears if we introduce non relativistic diffeomorphism invariance in a systematic manner. This is done by {\bf{gauging the global symmetry of the dynamical model in question}}{\footnote{Note the difference between gauging the symmetry and gauging the algebra \cite{BM4}}}. A general algorithm is developed to couple a Galileo symmetic model with background curvature \cite{BMM2, BM4, BMM1, BMM3, AM}. This theory, which has been inspired by gauging the Poincare symmetry (Poincare gauge theory) \cite{Utiyama, Kibble, Sciama}, may aptly be called Galilean gauge theory (GGT).
  
%  NRDI is an active area of research and there are various other approaches \cite{SW, ABP, BR, B1}. Results from Galilean gauge theory agree with many existing findings, whereas there are important exceptions. Thus it is desirable to investgate how the different approaches compare. The fulcrum of the line between the different approaches is how to couple an action with Galilean invariance  with back ground curvature. The need for this spatial NRDI arose from the work of Son and Wingate concerning the study of electron motion in strongly coupled three dimensional electron system \cite{SW}. However  action of \cite{SW} could only be shown to be invariant under the most general time dependent diffeomorphism, if anomalous transformation laws for the gauge fields is to be assumed. Not only that, to show galilean boost invariance in flat limit, one has to assume a certain condition between the boost and the gauge parameter. This is not only unnatural, but more significanly reduces the symmetries of the system. The action derived from GGT \cite{BMM3} contain one more term proportional to a geometric field than in \cite{SW}. It has been shown \cite{BMM3,BM4} that
%this term is responsible for invariane of the action under non-relativistic general coordinate transformations with fields transforming canonically whereas \cite{SW} can achieve it with anomalous field transformations. Not only that, the flat limit is easy and canonical in GGT while in  \cite{SW} it is problematic. With hindsight we can see that the problem lies in coupling the theory in NC manifold.

In this paper we will present a thorough analysis of the torsional NC geometry applying GGT. This is a new application where the connection is taken to be general. We will show how metric compatibility can be used to express the general connection in terms of the Dautcourt's symmetric form plus a part which resembles
the contortion tensor in Riemann Cartan space. Thus our analysis is able to pass to the known results \cite{BMM2} for symmetric connection. This is a unique feature of the GGT based analysis. We will also express the torsion tensor as a mixed form, where the part
containing the first order variables reveal the arbitrariness of the torsion
as the effect of the arbitrariness of the spin connection. The temporal part agrees with the usual result \cite{C1, GS, ABP1}.

Though our emphasis is on the torsional NC geometry, it will be advantagous to begin with the coupling of a Galilean invariant free particle theory with general NC manifold. This problem was earlier solved by taking the corresponding relativistic model, then coupling it with Riemann Cartan background and taking the $c \to \infty$ limit \cite{B1}. We find that using GGT it is practically one step job. In the process we will identify the torsional NC elements from GGT. Apart from fixing our notation, this will provide remarkable spin - offs,
as we will see.

The lay out of the present paper will now be described. In section 2
we provide an overview of the torsional NC connection, identifying some of the conflicting aspects of the existing results. The next section illustrates the Galilean gauge theory algorithm \cite{BMM1, BMM3, BM4} by taking the nonrelativistic free particle as an example. The fields introduced in the process are connected with general NC geometric structures
in this section, following the GGT approach \cite{BMM2}. This enables us to get an action for the free particle  coupled with background NC geometry. This is a new result from Galilean gauge theory, which was derived earlier from a relativistic theory \cite{B1} by taking the $c \to \infty$ limit. Moeover, this  gives an independent verification of the corresponding spatially diffeomorphic theory given by GGT \cite{BMM1}. A complete expression of the connection is given in terms of the metric properties and a combination of the torsion tensors, in section 3. This is the central result of the paper which shows the parallel to the Riemann Cartan manifold. Note that the combination of torsion tensors is an analogue of the contorsion tensor in Riemann Cartan geometry . An implicit mixed form of the torsion tensors is obtained where the firsi order variables appear in a suggestive manner. The temporal part of the torsion tensor can easily be isolated. This temporal part reproduces the well-known expression quoted in the literature. A list of the results
is provided. The paper ends with concluding remarks in section 5.
\section{Overview of the problem}
Non-relativistic diffeomorphism invariance (NRDI) has a long history.Cartan first demonstrated that the Newtonian theory of gravity can also be formulated as a geometric theory in a four dimensional manifold which is now called Newton Cartan (NC) spacetime.  The NC spacetime is however characterised by two different degenerate metrics. They follow from the $c \to \infty$ limit of the Riemannian metric. In these theories there is no torsion.
 In fact existence of torsion along with metric compatibility was argued to be accompanied with non integrability problem \cite{K}.
 When gravity was the focal theme, a symmetric connection proved to be sufficient. But  recent analysis requires torsional NC manifold to which field theories symmetric under Galilean transformations are to be coupled.
 
 However, the understanding of torsional connection is not still very clear. To see the point, it will be sufficient to consider a popular form of torsional connection given in \cite{GS}
 \begin{align}
{\Gamma^{ \sigma}}_{\mu\nu} & = \tau^{\sigma}\partial_{\nu}\tau_{\mu} +
\frac{1}{2}h^{\sigma\rho} \Bigl(\partial_{\nu}h_{\rho\mu} + \partial_{\mu}h_{\rho\nu} - \partial_{\rho}h_{\mu\nu}\Bigr) 
\label{ncm31}
\end{align}
Here $\{\tau^{\sigma}, h^{\sigma\rho},{\Gamma^{ \sigma}}_{\mu\nu}\} $
define a torsional NC manifold. %The connection ${\Gamma^{ \sigma}}_{\mu\nu}$ is claimed to be metric compatible.
 Similar connections have been prescribed in \cite{B1}, \cite{C1}. 

We define the torsion tensor as,
\begin{eqnarray}
T^\rho{}_{\mu\nu} = \Gamma_{\nu\mu}^{\rho} - \Gamma_{\mu\nu}^{\rho}\label{T}
\end{eqnarray}
This definition is in conformity with our definition of the covariant derivative
in the metric form
\begin{eqnarray}
\nabla_\mu A_\nu = \partial_\mu A_\nu - \Gamma^\rho_{\nu\mu} A_\rho\label{compat}
\end{eqnarray}
Using (\ref{ncm31}), we get 
\begin{equation}
T^\rho{}_{\mu\nu} =\tau^{\rho}\left(\partial_{\mu}\tau_{\nu}- \partial_{\nu}
\tau_{\mu}\right)
\end{equation}
The torsion tensor is purely temporal. The question arises why torsion should be purely temporal? In this context it may be remembered that \cite{ABP1},starting from gauging the algebra approach, has reported a spatial part of the torsion.

The second observation is the form of the connection. In torsionless geometry, the connection is given by the symmetric Dautcourt form
\begin{align}
{\Gamma^{D \sigma}}_{\mu\nu} & = \tau^{\sigma}\partial_{(\mu}\tau_{\nu)} +
\frac{1}{2}h^{\sigma\rho} \Bigl(\partial_{\nu}h_{\rho\mu} + \partial_{\mu}h_{\rho\nu} - \partial_{\rho}h_{\mu\nu}\Bigr)
\label{ncm21}
\end{align}
where, a term containing an arbitrary spatial shift is dropped, in order to make the comparision with (\ref{ncm31}) transparent.
The connection(\ref{ncm31}) can be written as 
\begin{equation}
{\Gamma^{\sigma}}_{\mu\nu} = {\Gamma^{D \sigma}}_{\mu\nu} -\frac{1}{2} T^\sigma{}_{\mu\nu}
\end{equation}
where ${\Gamma^{D \sigma}}_{\mu\nu}$ is symmetric and $T^\sigma{}_{\mu\nu}$ is antisymmetric under the exchange of $\mu$ and $\nu$.  We find that the symmetric part of the torsional connection is just the Dautcourt term, i.e. same as the torsionless connection. One becomes uneasy, for the geodesic motion given by the well known formula
 \begin{eqnarray}
\frac{d^2x^\mu}{d\lambda^2} + \Gamma^\mu_{\rho\nu}\frac{dx^\rho}{d\lambda}\frac{dx^\nu}{d\lambda} = 0
\end{eqnarray}
indicates that the geometry of the free fall is not affected by torsion.
In fact in Riemann Cartan space time with metric compatibility, the connection is given by
\begin{align}
{\Gamma^{ \sigma}}_{\mu\nu} & = 
\frac{1}{2}g^{\sigma\rho} \Bigl(\partial_{\nu}g_{\rho\mu} + \partial_{\mu}g_{\rho\nu} - \partial_{\rho}g_{\mu\nu}\Bigr) + C^\sigma_{\mu\nu}
\label{ncm22}
\end{align}
where
\begin{equation}
C^\sigma_{\mu\nu} = -\frac{1}{2}\left( T^\sigma{}_{\mu\nu} - T_\mu{}^\sigma{}_\nu + T_{\mu\nu}{}^\sigma\right)\label{rccon}
\end{equation}
is the contorsion tensor. The symmetric part of the connection (\ref{ncm22}) is given by 
\begin{align}
{\Gamma^{ \sigma}}_{\mu\nu} & = 
\frac{1}{2}g^{\sigma\rho} \Bigl(\partial_{\nu}g_{\rho\mu} + \partial_{\mu}g_{\rho\nu} - \partial_{\rho}g_{\mu\nu}\Bigr)-
\frac{1}{2}\left( T_{\mu\nu}{}^\sigma + T_{\nu\mu}{}^\sigma\right) 
\label{ncm222}
\end{align}
We observe that torsion do modify the symmetric part of the
connection (\ref{ncm22}), as shown by (\ref{ncm222}).
The NC manifold with torsion is, in principle obtainable as $c\to\infty$ limit of the Riemann Cartan manifold. Thus the torsional connection should be of a similar form as (\ref{ncm222}), where combination of torsion tensor contributes to the symmetric part.

We have discussed some results concerning torsional connection
available in the literature \cite{GS, C1, B1}. This is, however, sufficient to convince one self that the concept of torsion in NC geometry is plagued with riddles. The reason is not difficult to find. Torsion is best studied by the first order vielbein approach.
 An appropriate first order formulation is needed; one about which an unanimity is yet to be achieved \cite{ABP1, Son,BMM1, C1, GPR} .
% Though several authors have approached the problem , a clear picture did not emerge. Rather, certain confusions prevail. The study of the torsional NC geometry from different angles is therefore essential.

 The difficulty in the introduction of torsion in NC geomtry may be understood by comparing with Riemann-Cartan spacetime which is the torsionful manifold invariant under relativistic diffeomorphism .  %The spacetime in special relativity is $3 + 1$ dimensional Minkowski spacetime which is isometric under the Poincare group of transformations.
%  Gravity is formulated as a spacetime effect 
  %in General Relativity (GR) 
 In  Riemann-Cartan spacetime, metric is  factorisable in the vielbines. On the contrary, the NC spacetime is characterized by two degenerate metrics. So a direct approach to the vielbein formulation is problematic.
  One can of course introduce vielbeins and spin connections {\bf{geometrically}} but such diffeomorphic spacetime manifold is vaccuous until we define the physics in the tangent space \cite{W1}. In case of the relativistic manifold both space and time are relative.  Physical laws operating locally are the laws of special theory of relativity. Since space and time are both relative, the Riemann-Cartan spacetime can be foliated arbitrarily and local time direction can be defined normal to the spacelike leaves of the foliation. The situation in non relativistic diffeomorphism is very difficult. The absolute concept of Newtonian time is manifested in the fact that there is a unique foliation of the NC spacetime where Euclidean geometry holds locally in space with absolute time marching on \cite{K} \footnote{Globally this spatial slice may be curved}. In a given construction, one can fix this foliation for once. The corresponding coordinates are called the adapted coordinates. The properties of this foliation are  dictated by {\bf{the Galilean dynamics}}. Any arbitrary foliation will not give Galilean symmetry in the locally flat limit and the vielbein formalism will fail to exhibit  dual aspects of physical symmetries, namely local galilean symmetry and nonrelativstic diffeomorphism invariance.
  
  Galilean gauge theory approach \cite{BMM1, BMM2, BMM3, BM4} is just the right one to reach the Galilean frame by its construction.
Thus it can tackle the problems of NRDI in a straightforward way. This is no where more apparent than in discussing the torsional NC spacetime, as we will see in the following.
\section{Galilean gauge theory and Newton Cartan spacetime} 
 Galilean gauge theory as well as its connection with NC geometric elements have reached a more or less well developed stage \cite{BMM1, BMM2, BMM3, BM4}
 over the last one and half year. It is thus unnecessary to give a general review of the method.
On the other hand, an illustration of the algorithm with the help of a Galilean symmetric model as an example will be appropriate. This will help us to fix the notation and also provide new insights. We choose the example of a free particle moving nonrelativistically. The coupling of such
a theory with NC background will emerge naturally as a consequence of the algorithm.This is a new result in the context of GGT and reproduces earlier results obtained by taking nonrelativisctic limit of a relativistic theory \cite{B1}.

We consider the Schrodinger field action
\begin{equation}
S = \int dt  \int d^3x  \left[ \frac{i}{2}\left( \psi^{*}\partial_{0}\psi-
\psi\partial_0\psi^{*}\right) -\frac{1}{2m}\partial_k
\psi^{*}\partial_k
\psi\right].
\label{globalaction} 
\end{equation}
which is invariant under the Galilean transformations,
 \begin{equation}
  x^\mu \to x^\mu + \xi^\mu\label{globalgalileans}
\end{equation}
where $\xi^\mu$ is given by $\xi^{0} =-\epsilon$ and $\xi^{i} = \eta^{i}-v^{i}t $, 
   with
%\begin{equation}
$\eta^i=\epsilon^{i}+ \lambda^{i}{}_{j}x^{j}$
%\end{equation} 
 The constant parameters $\epsilon$, $\epsilon^{i}$, $\lambda^{ij}$ and $v^{i}$  respectively represent time and space translation, spatial rotations and galilean boosts.
 $\lambda^{ij}$ are antisymmetric under interchange of the indices. 
 If these parameters are elevated to functions of space and time, the corresponding symmetry will involve local Galilean transformations.
 
  The first step is to write the corresponding locally Galilean symmetric model in terms of local coordinates. One can readily write this, following the algorithm given in \cite{BMM1, BMM3, BM4}, as
\begin{equation}
S= \int dx^0d^3x\frac{\det(\Sigma_a{}^k)^{-1}}{\Sigma_0{}^0}\left[\frac{i}{2}(\psi^{*}\nabla_{0}\psi
-\psi\nabla_{{0}}\psi^{*}) -\frac{1}{2m}\nabla_a\psi^{*}
\nabla_a\psi
\right]
\label{localschaction} 
\end{equation} 
 The time coordinate in the local system will be denoted by ${0}$ and the space coordinates by $a$. Collectively, the local coordinates will be denoted by the initial letters of the Greek alphabet (i.e. $\alpha,\beta$ etc.).
  In order to restore the invariance of the theory (\ref{localschaction}) under local galilean transformations, we have replaced
ordinary derivatives by the local covariant derivatives
 \begin{eqnarray}
\nabla_{{0}}\psi &=&{\Sigma_{{0}}}^0D_0 \psi+{\Sigma_{{0}}}^k D_k\psi
\nonumber\\
\nabla_a\psi &=&{\Sigma_a}^{k}D_k\psi.
\label{nab1}
\end{eqnarray}
where, 
\begin{eqnarray}
D_k\psi=\partial_k\psi+iB_k\psi\nonumber\\
D_0\psi=\partial_t\psi+iB_0\psi \label{firstcov1}
\end{eqnarray}
are the covariant drivatives in the global coordinates.

The new fields $B_\mu$ and ${\Sigma_\alpha}^{\mu}$ introduced above have crucial physical significance.The gauge fields $B_0$ and $B_k$ correspond to gauging the rotations and galilean boosts.
They have the structures,
\begin{eqnarray}
B_k = \frac{1}{2}B_k^{ab}\omega_{ab} + B_k^{a{0}}\omega_{a}\nonumber\\
B_0 = \frac{1}{2}B_0^{ab}\omega_{ab} + B_0^{a{0}}\omega_{a}
\label{gaugefields}
\end{eqnarray}
where $\omega_{ab}$ and $\omega_{a}$ are respectively the generators of rotations and Galileo boosts. %Since we consider scalar fields only so $\lambda_{ab} = 0$. We may ignore $B_k^{ab}$ and $B_t^{ab}$
%However an important exception occurs in two space dimensions where the rotation generator is a scalar. This allows coupling with a scalar field, a fact crucial in the study of the FQHE as we will see. For the sake of generality we will keep the form ({\ref{gaugefields}}).
The fields ${\Sigma_\alpha}^{\mu}$, 'rotate' the global covariant derivatives (\ref{firstcov1}) into the local covariant drivatives (\ref{nab1}). The required transformations of  $\Sigma_\alpha{}^\mu$ are given by \cite{BM4},
\begin{align}
\delta_0 {\Sigma_0}^{k} &= -\xi^\nu {\partial_\nu{\Sigma}_0}^{k}+ {\Sigma_0}^{\nu}\partial_{\nu}\xi^{k} - v^b{\Sigma_b}^{k}\nonumber\\
\delta_0 {\Sigma_a}^{k} &= -\xi^\nu {\partial_\nu{\Sigma}_a}^{k}+ {\Sigma_a}^{\nu}\partial_{\nu}\xi^{k} -\lambda_a{}^b{\Sigma_b}^{k}
\label{delth3}
\end{align}
These transformations show that the lower index of $\Sigma_\alpha{}^\mu$ 
transform under the Galilean transformations, whereas the upper index transform under diffeomorphism,
$x^\mu \to x^\mu + \xi^\mu$. This observation allows one to think (\ref{localschaction}) as a theory in 4-dim curved space time.

It is also possible to define the curvature tensor by adopting the same path followed in Poincare gauge theory. We can define a field
tensor from the commutator of the covariant derivatives given in (\ref{nab1}). The curvature tensor may be abstracted from this field tensor. Of course nontrivial difficulty may arise due to the peculiarity of the nonrelativistic aspects. This may be pursued as an independent work but in the present paper we are studying geometrical interpretation, for which dynamics of the gauge fields is not necessary. 

The geometric interpretation run as follows.
The action (\ref{localschaction}) can be interpreted as a field thory in curved NC background.
The local coordinates $x^{\alpha}$ label the tangent space and the global coordinates $x^{\mu}$ chart the background curved spacetime.
The $\Sigma$ fields are the vielbein with $\Lambda$ as their inverse. They satisfy
\begin{equation}
{\Sigma_\alpha}^{\mu}\Lambda_{\mu}{}^{\beta}=\delta^{\beta}_{\alpha},~~~
{\Sigma_\alpha}^{\mu}\Lambda_{\nu}{}^{\alpha}=\delta^{\mu}_{\nu}
\label{sila}
\end{equation}
The spatial part $\Lambda_k{}^a$ is the inverse of $\Sigma_a{}^k$. %\cite{BMM2, BM4}.

 It has been proved that the 4-dim space referred above is NC
manifold.
 We begin with the definitions of the  metric \cite{BMM2}, 
 \begin{equation}
h^{\mu\nu}={\Sigma_a}^{\mu}{\Sigma_a}^{\nu}
\label{spm}
\end{equation}
and the one form
\begin{equation}
\tau_{\mu}={\Lambda_\mu}^{0}
\label{tem}
\end{equation}
where ${\Lambda_\mu}^{\alpha}$ is defined  in (\ref{sila}). In our coordinate system $\tau^\mu = \Sigma_0{}^0,\Sigma_0{}^k $. Also, since $\Sigma_a{}^0 = 0$,
\begin{eqnarray}
h^{\mu 0} = \Sigma_a{}^\mu\Sigma_a{}^0 = h^{0\mu} =0\label{me}
\end{eqnarray}
 Using the transformations (\ref{delth3}) and equation (\ref{sila}), we can show that both $h^{\mu\nu}$ and $\tau_\mu$ satisfy appropriate trnsformations under diffeomorhism of the manifold.

From the above definition we get
\begin{equation}
h^{\mu\nu}\tau_{\nu} = 0
\label{deg}
\end{equation}
which shows that the metric ${h^{\mu\nu}}$ is degenerate. We can, nevertheless, define a timelike vector $\tau^\mu$
as 
\begin{equation}
\tau^{\mu}=
{\Sigma_0}^{\mu}\hspace{.3cm};\hspace{.3cm}\tau_{\mu}\tau^{\mu} = 1
\label{tem1}
\end{equation}
and the covariant tensor
\begin{equation}
h_{\nu\rho}=\Lambda_{\nu}{}^{a} \Lambda_{\rho}{}^{a}
\label{spm2}
\end{equation}.
Clearly,
\begin{align}
h_{\mu\nu}\tau^\nu &=  {\Lambda_{\mu}}^a {\Lambda_\nu}^a {\Sigma_0}^\nu\notag\\ &=  {\Lambda_{\mu}}^a\delta_0^a\notag\\ &=0\label{n}
\end{align}
Finally %we can easily verify that 
\begin{equation}
h^{\mu\lambda}h_{\lambda\nu} = \delta^\mu_\nu - \tau^\mu\tau_\nu\label{spm3}
\end{equation}
Thus
$h^{\mu\nu}$,
 $\tau_\nu$ define the elements
 of the NC geometry. It will be useful to understand the meaning of this geometry. For example, one may enquire about the definition of proper distance. In Newtonian mechanics time is absolute and such distance can only be meaningful if an event in NC manifold is spacelike. The element $\tau_\mu$ determines whether an event $\xi^\mu$ is timelike ($\sqrt{\tau_\mu\tau_\nu \xi^\mu\xi^\nu} > 0$) or spacelike ($\sqrt{\tau_\mu\tau_\nu \xi^\mu\xi^\nu} = 0$). If the event is spacelike, one can define a distance unambiguously. The distance is defined as $\sqrt{h^{\mu\nu}\lambda_\mu\lambda_\nu}$, where $\xi^\mu = h^{\mu\nu}\lambda_\nu$ \cite{JC}. Interestingly, the Galilean gauge theory naturally arrives at the coordinate system where the direction of time flow matches with the absolute time.
 Here, events on the foliation have unique distance where time is not involved. This can be verified from the metric elements given in (\ref{me}).

It is now straightforward to verify that the action (\ref{localschaction} is equivalent to
\begin{equation}
S = \int dx^{0}d^3x \det{\Lambda_\mu}^\nu\left[\frac{i}{2}\tau^\mu\left(\psi^*D_\mu\psi -\psi D_\mu\psi^*\right) - \frac{1}{2m}h^{\mu\nu}D_\mu\psi^*D_\nu\psi\right]\label{J}
\end{equation}

The factor $dx^{0}d^3x \det
{\Lambda_
\mu}^\alpha$ is the volume form defined in the NC manifold. The above action is the cherished form of coupling of ({\ref{globalaction}}) with NC geometry. This result is in conformity with similar result derived from a relativistic theory by going to the $c\to \infty$ limit \cite{B1}.

 The action (\ref{J}) actually leads to yet another interesting finding.
 Note that the spatial coordinates in our calculation label the spacelike foliation which is unique
 in the NC spacetime whereas the time coordinate serves as the affine parameter ticking the absolute time.
 Specialising to our coordinates and putting the time translation to be zero, we recover the action invariant under most general nonrelativistic spatial diffeomorphism: $x^i \to x^i+ \xi^i$. It was earlier shown that $\Sigma_0{}^0 = const$ in this limit \cite{BMM3,BM4} which can be taken to be $1$ without any loss of generality.
 Denoting the determinant of the spatial metric $h_{kl}$ by $h$,
we can easily derive from definition (\ref{spm2}),
\begin{equation}
\det{\Lambda_\mu}^\nu =\Sigma_0{}^0 \det{\Lambda_i}^j= \sqrt{h}
\end{equation}

Now we can
write the action (\ref{J}) as, 
\begin{eqnarray}
S &=& \int dx^{0}d^3x \sqrt{h}\left[\frac{i}{2}\left(\psi^*D_0\psi -\psi D_{0}\psi^*\right) - \frac{1}{2m}h^{kl}D_k\psi^*D_l\psi\right. \nonumber\\&+& \left.\frac{i}{2}\Sigma_0{}^k\left(\psi^*D_k\psi -\psi D_k\psi^*\right)\right]
\label{J1}
\end{eqnarray}
In deriving the above we have made use of (\ref{tem1}).We find (\ref{J1}) is exactly what GGT predicts \cite{BMM1}.

The above action represents the coupling of nonrelativistic matter with  NC background metric. Note that the space time metric $h^{\mu\nu}$ is degenerate. So the usual procedure of defining the energy momentum tensor as a response to metric variation is not applicable here. Indeed it may be recalled that this definition of energy momentum tensor involves $\sqrt{g}$ in the denominator, where $g$ is the determinant of the metric $g_{\mu\nu}$ which vanishes here. Nevertheless, we can construct an energy momentum tensor by using Noether's prescription. This is possible  since we follow a field theoretic approach to develop our formalism.
 
In this connection it may also be mentioned that in the context of study of transport phenomena in Hall fluid, the action (\ref{globalaction}) was coupled minimally with curved background in  
\cite{SW}, supposedly diffeomorphic under arbitrary time dependent  spatial transformation. Their theory was (\ref{J1}) without the second term \footnote{Actually, in \cite {SW}, there was an external gauge field. But our arguments hold also in the more general case \cite{BMM3}}.
Naturally, their theory failed to be invariant under time dependent deformation. Invariance could only be achieved by assuming non canonical transformation for the fields. Also, as we see clearly here, they did not consider the spin connection. The result is, one has to assume a condition between the Galilean boost and the gauge parameter in order to retrive Galilean symmetry in the flat limit. Not only unnatural, such equality reduces symmetry of the system. 

% and not the one that would follow from \cite{SW}. Since (\ref{J1}) follows from (\ref{J}) which is again independently obtained in \cite{B1}, the GGT construction of spatially diffeomorphic theory is independently verified. %5\footnote{Note that the expression of $\tau^\mu$ and $h^{\mu\nu}$ is also obtained by gauging the extended Galilean algebra \cite{ABP}.}

\section{ Torsional Newton-Cartan geometry} In the above, the geometric connection (\ref{spm} - \ref{spm3}) was exploited to develop a first order (vielbein) formalism of the  Newton Cartan geometry. Thus via GGT we have been able to construct NC geometry. We have used these metric relations to couple a nonrelativistic free particle theory with the Newton Cartan background. We still left the issue of connection open. In the standard Newton Cartan theory, the connection is symmetric. Earlier, we have shown how to derive the standard NC connection from the Galilean gauge theory by imposing symmetry of the connection as an additional condition \cite{BMM2}. But there is nothing in our construction thst demands this symmetry. 
In this work we will withdraw the assumption of symmetric connecion and only demand
  metric compatibility. One then hopes that this analysis may naturally lead to the torsional NC geometry and help clarify the concept of torsion in such spaces.The following analysis shows that this hope will not be belied.
 
 We have already derived the metric structure of a general NC theory in terms of the first order theory variables. To proceed further we have to write the affine connection from our theory.
 The affine connection $\Gamma_{\nu\mu}^{\rho}$ in the first order form is easy to abstract from the vielbein postulate,
 \begin{equation}
\nabla_\mu{\Lambda^\alpha}_{\nu} = \partial_{\mu}{\Lambda^\alpha}_{\nu} - \Gamma_{\nu\mu}^{\rho}{\Lambda^\alpha}_{\rho}
+B^{\alpha}{}_{\mu\beta}{\Lambda^\beta}_{\nu} =0 
 \label{P}
\end{equation}
and is given by
 \begin{eqnarray}
\Gamma_{\nu\mu}^{\rho} &= \partial_{\mu}{\Lambda_{\nu}}^\alpha {\Sigma_\alpha}^{\rho}
+B^{\alpha}{}_{\mu\beta}{\Lambda_{\nu}}^\beta
{\Sigma_\alpha}^{\rho}\label{con1}
\end{eqnarray}
where the spin connection $B^{\alpha}{}_{\mu\beta}$ is introduced in (\ref{gaugefields}).
%We can write
%We will see in our formalism torsion can be naturally introduced.
The connection (\ref{con1}) is demonstrated to be metric compatible in \cite{BMM2}. Note in this connection that in the torsionless case, the spin connection may be expressed in terms of the vielbeins \cite{ABP,BMM2}. However in the present torsional example, the spin connections are independent.
From this one can expect that torsion may not be expressible in terms of metric variables alone.

%Also the identifica} &= \partial_{\mu}{\Lambda_{\nu}}^\alpha {\Sigma_\alpha}^{\rho}
%+B^{\alpha}{}_{\mu\beta}{\Lambda_{\nu}}^\beta{\Sigma_\alpha}^{\rho}\label{con1}
%\end{eqnarray}
In the general case the connection does not have definite symmetry. However we can always split $\Gamma_{\nu\mu}^{\rho}$ in its symmetric (S) and antisymmetric (A) parts,
\begin{eqnarray}
\Gamma_{\nu\mu}^{\rho} = {\Gamma^S}_{\nu\mu}^{\rho} + {\Gamma^A}_{\nu\mu}^{\rho}\label{split}
\end{eqnarray}
where,$ {\Gamma^A}_{\nu\mu}^{\rho}$ and $ {\Gamma^S}_{\nu\mu}^{\rho}$ are, respectively, given by
\begin{align}
{\Gamma^A}_{\nu\mu}^{\rho}
&=\frac{1}{2}[\partial_{\mu}{\Lambda_{\nu}}^\alpha {\Sigma_\alpha}^{\rho}-\partial_{\nu}{\Lambda_{\mu}}^\alpha {\Sigma_\alpha}^{\rho}+ B^{\alpha}{}_{\mu\beta}{\Lambda_{\nu}}^\beta{\Sigma_\alpha}^{\rho}-B^{\alpha}{}_{\nu\beta}{\Lambda_{\mu}}^\beta{\Sigma_\alpha}^{\rho}]\notag\\ &=\frac{1}{2}[\partial_{\mu}{\Lambda_{\nu}}^{0} {\Sigma_0}^{\rho}-\partial_{\nu}{\Lambda_{\mu}}^0 {\Sigma_0}^{\rho}+\partial_{\mu}{\Lambda_{\nu}}^a {\Sigma_a}^{\rho}-\partial_{\nu}{\Lambda_{\mu}}^a {\Sigma_a}^{\rho}\notag\\ &+B^{a}{}_{\mu 0}{\Lambda_{\nu}}^{0}{\Sigma_a}^{\rho}-B^{a}{}_{\nu 0}{\Lambda_{\mu}}^0
{\Sigma_a}^{\rho}+B^{a}{}_{\mu b}{\Lambda_{\nu}}^b{\Sigma_a}^{\rho}-B^{a}{}_{\nu b}{\Lambda_{\mu}}^b{\Sigma_a}^{\rho}]\notag\\
 \label{PPA}
\end{align}
and
\begin{align}
{\Gamma^S}_{\nu\mu}^{\rho}
&=\frac{1}{2}[\partial_{\mu}{\Lambda_{\nu}}^\alpha {\Sigma_\alpha}^{\rho}+\partial_{\nu}{\Lambda_{\mu}}^\alpha {\Sigma_\alpha}^{\rho}+ B^{\alpha}{}_{\mu\beta}{\Lambda_{\nu}}^\beta{\Sigma_\alpha}^{\rho}+B^{\alpha}{}_{\nu\beta}{\Lambda_{\mu}}^\beta{\Sigma_\alpha}^{\rho}]\notag\\ &=\frac{1}{2}[\partial_{\mu}{\Lambda_{\nu}}^{0} {\Sigma_0}^{\rho}+\partial_{\nu}{\Lambda_{\mu}}^0 {\Sigma_0}^{\rho}+\partial_{\mu}{\Lambda_{\nu}}^a {\Sigma_a}^{\rho}+\partial_{\nu}{\Lambda_{\mu}}^a {\Sigma_a}^{\rho}\notag\\ &+B^{a}{}_{\mu 0}{\Lambda_{\nu}}^{0}{\Sigma_a}^{\rho}+B^{a}{}_{\nu 0}{\Lambda_{\mu}}^0{\Sigma_a}^{\rho}+B^{a}{}_{\mu b}{\Lambda_{\nu}}^b{\Sigma_a}^{\rho}+B^{a}{}_{\nu b}{\Lambda_{\mu}}^b{\Sigma_a}^{\rho}]\notag\\
 \label{PPS}
\end{align}

The form of the symmetric part in terms of the NC structures will now be worked out. Using (\ref{tem})in (\ref{PPS}) we obtain
\begin{align}
{\Gamma^S}_{\nu\mu}^{\rho}
 &=\tau^{\rho}\partial_{(\mu}\tau_{\nu)} + \frac{1}{2}[\partial_{\mu}{\Lambda_{\nu}}^a {\Sigma_a}^{\rho}+\partial_{\nu}{\Lambda_{\mu}}^a {\Sigma_a}^{\rho}\notag\\ &+B^{a}{}_{\mu 0}{\Lambda_{\nu}}^{0}{\Sigma_a}^{\rho}+B^{a}{}_{\nu 0}{\Lambda_{\mu}}^0{\Sigma_a}^{\rho}+B^{a}{}_{\mu b}{\Lambda_{\nu}}^b{\Sigma_\alpha}^{\rho}+B^{a}{}_{\nu b}{\Lambda_{\mu}}^b{\Sigma_a}^{\rho}]\notag\\
 \label{PPS1}
\end{align}
To proceed further, note that if we assume the connection to be symmetric, the corresponding theory is torsionless NC theory. This symmetric connection is entirely given by metric quantities \cite{K}. We will see how it is modified in presence of torsion. As far as we know, this analysis is a novel one in the literature.

 Using (\ref{PPA}) and (\ref{n}) we can write,
\begin{align}
h_{\lambda\rho}{T}_{\mu\nu}^{\rho}
 &=[\partial_{\mu}{\Lambda_{\nu}}^a {\Sigma_a}^{\rho}-\partial_{\nu}{\Lambda_{\mu}}^a {\Sigma_a}^{\rho}\notag\\ &+B^{a}{}_{\mu 0}{\Lambda_{\nu}}^{0}{\Sigma_a}^{\rho}-B^{a}{}_{\nu 0}{\Lambda_{\mu}}^0{\Sigma_a}^{\rho}+B^{a}{}_{\mu b}{\Lambda_{\nu}}^b{\Sigma_a}^{\rho}-B^{a}{}_{\nu b}{\Lambda_{\mu}}^b{\Sigma_a}^{\rho}]h_{\lambda\rho}\notag\\
 \label{PPA1}
\end{align}
where the definition (\ref{T}) of the torsion has been used. Now, $\Sigma_a{}^\rho h_{\lambda\rho} = \Lambda_{\lambda a}$. Equation (\ref{PPA1})thus becomes
\begin{eqnarray}
\partial_{\mu}{\Lambda_{\nu}}^a \Lambda_{\lambda a}&-&\partial_{\nu}{\Lambda_{\mu}}^a \Lambda_{\lambda a} + B^{a}{}_{\mu 0}{\Lambda_{\nu}}^{0}\Lambda_{\lambda a}-B^{a}{}_{\nu 0}\Lambda_{\mu}^0\Lambda_{\lambda a}\nonumber\\&+& B^{a}{}_{\mu b}{\Lambda_{\nu}}^b\Lambda_{\lambda a}-B^{a}{}_{\nu b}{\Lambda_{\mu}}^b\Lambda_{\lambda a} = h_{\lambda\rho}{T}_{\mu\nu}^{\rho}
\end{eqnarray}
Permuting the indices $\mu, \nu, \lambda $ cyclically twice and subtracting the first outcome from the second, we get,
\begin{eqnarray}
B^{a}{}_{\mu b}{\Lambda_{\nu}}^a\Lambda_{\lambda b}& +& B^{a}{}_{\nu a}{\Lambda_{\mu}}^a\Lambda_{\lambda b}
= -\partial_\nu\Lambda_{\lambda a}{\Lambda_{\mu}}^a  +\partial_\lambda h_{\mu\nu} -\partial_\mu \Lambda_{\lambda a}{\Lambda_{\nu}}^a \nonumber\\&-& \left[\left(B_\mu{}^{a0}\tau_
\lambda - B_\lambda{}^{a0}\tau_\mu\right)\Lambda_{\nu
a} -\left(B_\nu{}^{a0}\tau_\lambda - B_\lambda{}^{a0}\tau_\nu\right)\Lambda_{\mu
a} \right]\nonumber\\&+& h_{\mu\rho}{T}_{\nu\lambda}^{\rho} -  h_{\nu\rho}{T}_{\lambda\mu}^{\rho}\label{b}
\end{eqnarray}
Substituting (\ref{b}) in (\ref{PPS1}) we get the desired expression for the symmetric part of the connection,
%has already been worked out in \cite{BMM2, BM4}. There we have seen that the expression becomes 
\begin{align}
{\Gamma^{S\sigma}}_{\mu\nu} & = \tau^{\sigma}\partial_{(\mu}\tau_{\nu)} +
\frac{1}{2}h^{\sigma\rho} \Bigl(\partial_{\nu}h_{\rho\mu} + \partial_{\mu}h_{\rho\nu} - \partial_{\rho}h_{\mu\nu}\Bigr) +  h^{\sigma\lambda}K_{\lambda(\mu}\tau_{\nu)}
\notag\\ &- \frac{1}{2}h^{\sigma\lambda}\left[ h_{\mu\rho}{T}_{\lambda\nu}^{\rho} + h_{\nu\rho}{T}_{\lambda\mu}^{\rho}\right]
\label{ncm1}
\end{align}
where, the penultimate term consists of an arbitrary shift by the two form K,
\begin{eqnarray}
 K_{\lambda\mu} = [B_{\mu}{}^{a0}{\Lambda_{\lambda a}} - B_{\lambda}{}^{a0}{\Lambda_{\mu a}}]\label{K}
\end{eqnarray}
Expression (\ref{ncm1}) is the symmetric part of the connection 
%( Dautcourt's formula) 
for the torsional NC spacetime. If we additionaly impose the  symmetry
\begin{equation}
\Gamma^\rho_{\mu\nu} = \Gamma^\rho_{\nu\mu}
\end{equation}
the connection becomes
\begin{align}
{\Gamma^{D \sigma}}_{\mu\nu} & = \tau^{\sigma}\partial_{(\mu}\tau_{\nu)} +
\frac{1}{2}h^{\sigma\rho} \Bigl(\partial_{\nu}h_{\rho\mu} + \partial_{\mu}h_{\rho\nu} - \partial_{\rho}h_{\mu\nu}\Bigr) +  h^{\rho\lambda}K_{\lambda(\mu}\tau_{\nu)}
\label{ncm2}
\end{align}
which is Dautcourt's formula for the connection in standard (torsionless) NC spacetime.

Coming back to the torsional NC geometry we can easily see that the general connection can be split as
\begin{eqnarray}
\Gamma^\rho_{\mu\nu} = {\Gamma^{D \sigma}}_{\mu\nu} + C^{D \sigma}{}_{\mu\nu} 
\label{split1}
\end{eqnarray}
where, 
\begin{eqnarray}
C^{D \rho}{}_{\mu\nu} = -\frac{1}{2}\left[{T}_{\mu\nu}
^{\rho} -h^{\rho\lambda} h_{\mu\sigma}{T}_{\lambda\nu}^{\sigma} + h^{\rho\lambda} h_{\nu\sigma}{T}_{\mu\lambda}^{\sigma}\right]
\label{C}
\end{eqnarray}
Note that $C^{D \sigma}{}_{\mu\nu}$
has no definite symmetry in conformity to the fact that connection is of the most general type. 

Equation (\ref{split1}) is one of the most significant results of the present report.
We see that the symmetric part of the connection (\ref{ncm1}) contain two parts. The first one is the Dautcourt connection, the appropriate connection for the torsionless case. The second part originates due to torsion. In torsional NC geometry, the symmetric part of the connection
 is influenced by the torsion. This influence has no temporal part because, quite generally, identity (\ref{spm3}) enforces it in the spatial sector. The definition of straight line as a line parallel to itself is thus influenced by the torsion. This is expected because in Riemann Cartan space similar phenomenon exists, the difference lies in the fact that the latter is not confined to spatial sector. The similarity and the difference are both expected. Torsional NC space time is in principle obtainable as $c \to \infty$ limit of the Riemann Cartan space. The difference is explained by the existence of absolute time in NC geometry.

 Secondly, a scrutinee of the result shows that there is an arbitrary two form in the expression of the symmetric connection, just like the torsionless NC theory. In fact the torsionless theory
should be recovered from the more general theory by putting torsion to be zero. Surprisingly this requirement is not much emphasised in the literature. 
 
%There is however one point regarding the derivation of (\ref{conec}). In \cite{BMM2}
%we have used ${\Gamma^A}_{\mu\nu}^{\rho} = 0$. In the general case this is not true.
%If we replace the antisymmetric part situation will not change because after getting the final result we can again symmetrize whereby the antisymmetric correction vanished. The expression for the symmetric part remains to be given by (\ref{conec}).

Our next task is to express the torsion (\ref{T}). Primarily, we get from (\ref{PPA}) that
\begin{align}
{T}_{\mu\nu}^{\rho}
&=\tau^\rho\left(\partial_{\mu}{\tau_{\nu}} -\partial_{\nu}{tau_{\mu}}\right) +{\Sigma_a}^{\rho}[\partial_{\mu}{\Lambda_{\nu}}^a -\partial_{\nu}{\Lambda_{\mu}}^a \notag\\ &+B^{a}{}_{\mu 0}{\Lambda_{\nu}}^{0}-B^{a}{}_{\nu 0}{\Lambda_{\mu}}^0
+B^{a}{}_{\mu b}{\Lambda_{\nu}}^b-B^{a}{}_{\nu b}{\Lambda_{\mu}}^b]
 \label{T1}
\end{align}
As one can see, the temporal part is
\begin{eqnarray}
\tau_\rho  T^\rho{}_{\mu\nu}
= \left(\partial_\mu\tau_\nu - \partial_\nu\tau_\mu\right)
\end{eqnarray}
as a consequence of
\begin{equation}
\tau_\rho{\Sigma_a}^{\rho}={\Lambda_\rho}^{0}{\Sigma_a}^{\rho}= \delta_a^0=0 
\end{equation}
The expression of the temporal part agrees universally. 

However (\ref{T1}) also shows that torsion is not purely temporal. The spatial part is given in (\ref{T1}) in the first order variables. Question is, can it be expressed entirely in terms of metric structures. Comparision with Riemann Cartan space once again gives us guideline. It suggests that torsion should be dependent on independent spin connection \cite{carroll}. Below, we will derive an implicit relation involving torsion, where this dependence is manifest.

 Starting from the definition (\ref{T}) we get after a few steps
\begin{eqnarray}
{T^\rho}_{\mu\nu} &=& \tau^\rho\left(\partial_\mu\tau_\nu - \partial_\nu\tau_\mu\right)\nonumber\\&+& h^{\rho\sigma}\Lambda_{\sigma a}\left[\left(\partial_\mu h_{\nu\lambda}\right)\Sigma_a{}^\lambda  + h_{\nu\lambda}\left(\partial_\mu\Sigma_a{}^\lambda
\right) -\left(\partial_\nu h_{\mu\lambda}\right)\Sigma_a{}^\lambda  - h_{\mu\lambda}\left(\partial_\nu\Sigma_a{}^\lambda\right)
 \right]\nonumber\\&+& h^{\rho\sigma}\Lambda_{\sigma a}\left[\left(B_\mu{}^{ab}\Lambda_{\nu b} -B_\nu{}^{ab}\Lambda_{\mu b}  \right) +\left(B_\mu{}^{a0}\Lambda_{\nu 0} -B_\nu{}^{a0}\Lambda_{\mu 0} \right) \right]\label{condt}
\end{eqnarray}

Now note that 
\begin{eqnarray}
\Sigma_a{}^\lambda\Lambda_{\sigma a} =\Sigma_\alpha{}^\lambda\Lambda_{\sigma \alpha}- \Sigma_0{}^\lambda\Lambda_{\sigma 0} = \delta^\lambda_\sigma - \tau_\sigma\tau^\lambda .\label{rel1}
\end{eqnarray}
 When multiplied by $h^{\rho\sigma}$, the result is
 \begin{equation}
 h^{\rho\sigma}\Sigma_a{}^\lambda\Lambda_{\sigma a}          = h^{\rho\lambda}\label{res1}
\end{equation} 
Contribution from the second term of the right hand side of (\ref{rel1}) vanishes. This result will be important in various simplifications.

Now, using (\ref{b}) we can express
% $ A=
%(B_\mu{}^{ab}\Lambda_{\nu a}\Lambda_{\sigma b} -B_\nu{}^{ab}\Lambda_{\mu a}\Lambda_{\sigma b})$ as
\begin{eqnarray}
 (B_\mu{}^{ab}\Lambda_{\nu a}\Lambda_{\sigma b} &-&B_\nu{}^{ab}\Lambda_{\mu a}\Lambda_{\sigma b})= \Lambda_\mu{}^a\partial_\sigma\Lambda_\nu{}^a -
\Lambda_\nu{}^a\partial_\sigma\Lambda_\mu{}^a
+ 2(\partial_\mu h_{\sigma\nu} - \partial_\nu h_{\sigma\mu}\nonumber\\& +&\partial_\nu\Lambda_{\sigma a}\Lambda_{\mu a}
-\partial_\mu\Lambda_{\sigma a}\Lambda_{\nu a}
\nonumber\\&+& [B_\mu{}^{a0}(\tau_\sigma\Lambda_{\nu a} +
\tau_\nu \Lambda_{\sigma a}) -  B_\mu{}^{a0}(\tau_\sigma\Lambda_{\nu a} +
\tau_\nu \Lambda_{\sigma a})]\nonumber\\& +&(h_{\nu\lambda} T^\lambda{}_{\sigma \mu} -h_{\mu\lambda} T^\lambda{}_{\sigma \nu} +
2h_{\sigma\lambda} T^\lambda{}_{\nu \mu} )\label{a}
\end{eqnarray}

Substituting (\ref{a}) and  (\ref{res1}) in  
(\ref{condt}) we obtain after some simplifications,
\begin{eqnarray}
T^\rho{}_{\mu\nu} &-& h^{\rho \sigma}h_{\nu\lambda}  T^\lambda{}_{\sigma\mu} + h^{\rho \sigma}h_{\mu\lambda} T^\lambda{}_{\sigma\nu}= \tau^\rho\left(\partial_\mu\tau_\nu - \partial_\nu\tau_\mu\right)
- h^{\rho\sigma}\left(\partial_\mu h_{\nu\sigma} - \partial_\nu h_{\mu\sigma}\right)\nonumber\\&+& h^{\rho\sigma}\left[ \Lambda_{\nu a}\partial_\sigma\Lambda_{\mu a} -\Lambda_{\mu a}\partial_\sigma\Lambda_{\nu a} + 2B_\sigma{}^{ab}\Lambda_{\nu a}\Lambda_{\mu b} \right]   \label{condtn}
\end{eqnarray}
 Note that first order variables occur in a remarkable combination,$\Lambda_{\nu a}\partial_\sigma\Lambda_{\mu a} -\Lambda_{\mu a}\partial_\sigma\Lambda_{\nu a} + 2B_\sigma{}^{ab}\Lambda_{\nu a}\Lambda_{\mu b}$. We can extend it to 
 \begin{eqnarray}
 \Lambda_{\nu a}\partial_\sigma\Lambda_{\mu a} -\Lambda_{\mu a}\partial_\sigma\Lambda_{\nu a} + 2B_\sigma{}^{ab}\Lambda_{\nu a}\Lambda_{\mu b}&+& B_\sigma{}^{a0}\Lambda_{\nu a} \Lambda_{\mu 0}-B_\sigma{}^{a0}\Lambda_{\mu a}\Lambda_{\nu 0}
\nonumber\\ &-&\left[B_\sigma{}^{a0}\Lambda_{\nu a} \Lambda_{\mu 0}-B_\sigma{}^{a0}\Lambda_{\mu a}\Lambda_{\nu 0}\right]
 \end{eqnarray}
 Now
 \begin{eqnarray}
 \Lambda_{\nu a}\partial_\sigma\Lambda_{\mu a} -\Lambda_{\mu a}\partial_\sigma\Lambda_{\nu a} + 2B_\sigma{}^{ab}\Lambda_{\nu a}\Lambda_{\mu b}&+& B_\sigma{}^{a0}\Lambda_{\nu a} \Lambda_{\mu 0}-B_\sigma{}^{a0}\Lambda_{\mu a}\Lambda_{\nu 0}\nonumber\\& =& \Lambda_\nu{}^aD_\sigma\Lambda_\mu{}^a -\Lambda_\mu{}^aD_\sigma\Lambda_\nu{}^a
\end{eqnarray}  
where, $D_\sigma\Lambda_\mu{}^a$ is the covariant derivative with respect to spin connection. Substitution of the above in (\ref{condtn}), we get
\begin{eqnarray}
T^\rho{}_{\mu\nu} &-& h^{\rho \sigma}h_{\nu\lambda}  T^\lambda{}_{\sigma\mu} + h^{\rho \sigma}h_{\mu\lambda} T^\lambda{}_{\sigma\nu}= \tau^\rho\left(\partial_\mu\tau_\nu - \partial_\nu\tau_\mu\right)- h^{\rho\sigma}\left(\partial_\mu h_{\nu\sigma} - \partial_\nu h_{\mu\sigma}\right)
\nonumber\\&+& h^{\rho \sigma}\left( \Lambda_\nu{}^aD_\sigma\Lambda_\mu{}^a -\Lambda_\mu{}^aD_\sigma\Lambda_\nu{}^a
\right) - h^{\rho \sigma}\left[\left(B_\sigma{}^{a0}\Lambda_{\nu a} \Lambda_{\mu 0}-B_\sigma{}^{a0}\Lambda_{\mu a}\Lambda_{\nu 0}\right)\right ]
   \label{condtnn}
\end{eqnarray}

In the following let us summarise the findings of the present paper:
\begin{enumerate}
\item The coupling of a nonrelativistic matter theory
with the general NC background has been achievd by GGT.

\item As a special case of the above
we have studied the corresponding spatially diffeomorphic action. We found that the prescription of GGT
is favoured, rather than that of \cite{SW}.

\item A full expression of the torsional connection is obtained as a sum of the Dautcourt form and a combination of the torsion which can be compared with the contorsion tensor in Riemann - Cartan spacetime.
 The symmetric part of it is modified by the contribution coming from this combination. 
\item It should be stressed that we followed the standard method of gauging for obtaining the torsion. In particular we have assigned compensating fields for time and space translation as well as rotations and boosts (see equation (\ref{gaugefields})). The nonrelativistic spin connection defined there is used to construct the connection (\ref{con1}). The torsion is just the antisymmetric part of this connection. The Dautcourt connection does not play any role in this derivation. The metric forms including the Dautcourt connection and the contorsion
have been invoked to trace a parallel between Riemann Cartan spacetime and the torsional NC spacetime.
\item The 'temporal' part of the torsion 
can  be identified, 
\begin{eqnarray}
\tau_\rho{T^\rho}_{\mu\nu} = \left(\partial_\mu\tau_\nu - \partial_\nu\tau_\mu\right)
\end{eqnarray}
This expression is well known in the literature \cite{ABP1,GS,C1}.
%the same as reported by nany authors \cite{C1, B1, D1, GA}. 

%\item A remarkable point is the appearance of the arbitrary two form $K_{\mu\nu}$ in the antisymmetrized combination in 
%(\ref{cont}). Since a symmetrized combination of $K_{\mu\nu}$ appears in Dautcourt's formula, a term involving $K_{\mu\nu}$ is expected in the expression of the torsion. However, to our knowledge this important aspect has so far alluded observation in the literature.
% \cite{ref1, ref2, ref3, ref 4}.

\item Our formulation is based on a systematic analysis of NC geometry. This is noteworthy, for sometimes the torsional connection has been assumed as an ansatz \cite{GS} where torsion is purely temporal. The theoretical premises of the ansatz  are not quite obvious . Often the deductions are quite intricate and there is no clear algorithm to understand the connections between the vielbein and metric formulations of the NC geometry \cite{ABP1}. On the contrary, the Galilean gauge theory offers a complete and natural formulation of the torsional Newton Cartan geometry, as has been shown here.
\end{enumerate}

\section{ Concluding remarks }
In the above analysis we have demonstrated how torsional Newton Cartan (NC)geometry can be obtained in Galilean gauge theory (GGT), a recent method of obtaining non relativistic diffeomorphism, formulated in \cite{BMM1} and applied successfully to a host of phenomena \cite{BMM2,BM4,BMM3,AM}. There are several approaches to the problem. However, there are differences of opinion in current literature \cite{ABP1,GS,C1} about various aspects of the first order formalism of NC geometry. We have, therefore,
derived the general metric structure from an example. This served to fix our notations. But remarkably, we get new results in the process. We started from  a nonrelativistic free particle theory
having Galilean symmetry in thr flat space. The algorithm of GGT was applied to eventually couple it with  NC background and one will hardly miss the relative ease with which the coupling was done. The structure of the coupled theory matched exactly with an earlier result \cite{B1}, obtained by the approach of taking the $c \to \infty$ limit of a relativistic model.

Anothr spin off is a resolution of the problem of exact nonrelativistic diffeomorphism invariant action for a particle. The problem is well known. From GGT we have earlier derived an action \cite{BMM1, BMM3, BM4} which had the correct symmetry dictated by {\it{appropriate transformations}} of the fieds under nonrelativistic diffeomorphism. The result from GGT differs from that of \cite{SW}, where however some of the fields have to transform anomalously
for the invariance to remain valid.
%From the coupled theory, we got the spatially diffeomorphic counterpart. 
The model proposed in \cite{SW} also did not have a proper flat limt -the gauge parameter gets mingled up with the boost parameter. This not only reduces symmetry but also makes the curved space coupling of a free theory impossible %\cite{BGM}
. In this paper we find that a proper coupling with a background curvature in space which follows from the corresponding theory in NC space time agrees with the form provided by GGT.

 As we have considered genral NC geometry, we found the general expression
 of the conection in terms of the metric variables and torsion. It has a clear division in terms of the well known Dautcourt form and a combination of the torsion tensor which formally resembles with the contortion tensor in Riemann Cartan space. From this expression (\ref{split1}) the passage to torsionless limit is crystal clear.
 
 We also obtained  expression for torsion. The temporal part of the torsion tensor agrees with the well known result
 in the literature. But our analysis clearly shows that it is not the whole story, There is a spatial part of the torsion tensor which is not entirely expressible in terms of the metric variables. This is also similar to relativistic theories. The metric variables were extracted as far as possible and an implicit relation for torsion \ref{condtnn} could be provided which explains our inability. The point is that in a general torsional manifold the turning of the orthogonal local frame contributes to the torsion.
 
 The construction of NC geometry with torsion is provided by using Galilean gauge theory. In a subject riddled with so many confusing points, our results stand in their own merit.% when geometrically interpreted, leads naturally to the unique foliation of the NC geometry where the coordinate time denotes the absolute time.
  As GGT is based on localising the symmetry of a dynamical theory, the essence of the Galilieo Newton concept of spacetime is inbuilt in it. Once again, we observe the merit of it in studying nonrelativistic diffeomorphism.

%\end{enumerate}

\end{document}